\documentclass[a4paper]{mem}
\usepackage{natbib}
\usepackage{graphicx}
\usepackage[a4paper]{hyperref}
\idline{73}{23}
\newcommand{\permil}{$^{o}$/$_{oo}$}
\newcommand{\msun}{\mbox{${\rm M}_{\odot}$ }}
\begin{document}

   \title{Constraints on AGB models \\ from the heavy-element composition \\ of 
presolar SiC grains 
}

   \author{M. Lugaro \inst{1}, 
   A. M. Davis \inst{2}, 
          R. Gallino \inst{3}, M. R. Savina \inst{5}, \and 
         M. J. Pellin \inst{5} \fnmsep
}

   \offprints{M. Lugaro}
\mail{Institute of Astronomy, University of Cambridge,
Madingley Rd, Cambridge CB3 0HA UK}

   \institute{Institute of Astronomy, University of Cambridge,
Madingley Rd, Cambridge CB3 0HA UK \email{mal@ast.cam.ac.uk}\\ 
              \and Enrico Fermi Institute, Department of the 
Geophysical Sciences, and Chicago Center for Cosmochemistry, University 
of Chicago, Chicago, IL 60637, USA\\
              \and Dipartimento di Fisica Generale, Universit\'a di Torino,
via P. Giuria 1, 10125 Torino, Italy\\ 
              \and Materials Science Division, Argonne National 
Laboratory, Argonne, IL 60439, USA
             }

   \abstract{

Presolar SiC grains formed around Asymptotic Giant Branch (AGB) stars during 
their carbon-rich phase and contain heavy elements in trace amounts showing the signature 
of the $slow$ neutron capture process ($s$ process). Thanks to recent advances in 
analysis techniques, SiC data now provide extremely precise information on neutron capture 
cross sections and AGB models. For example, high-precision data for Mo in single SiC grains 
indicate that a revision of the $^{95}$Mo neutron capture cross section is needed, while data for 
Zr indicates that the $^{22}$Ne($\alpha$,n)$^{25}$Mg reaction cannot be a dominant neutron
source for the $s$ process in AGB stars. We present model predictions for the
composition of Fe-peak elements in AGB stars. These elements could be analysed in the
near future thus providing further stringent constraints to our understanding of AGB
stars.

   }

   \authorrunning{M. Lugaro et al.}
   \titlerunning{Presolar SiC grains}
   \maketitle
%

\section{Introduction}

Among presolar meteoritic material, silicon carbide (SiC) grains have isotopic 
compositions very different to those commonly measured in the solar system. The majority 
of these grains are believed to have formed in the extended envelopes of carbon-rich 
Asymptotic Giant Branch (AGB) stars. A strong indication of this origin is the 
composition of elements heavier than Fe and present in very small amounts ($trace$), 
showing the unequivocal signature of the $slow$ neutron capture process ($s$ process) 
that occurs in AGB stars \citep{gallino:98,busso:01}.

The bulk of $s$-process elements in AGB stars are produced neutrons released by 
the $^{13}$C($\alpha$,n)$^{16}$O reaction. The $^{13}$C neutron source is activated 
in a tiny region in the top layers of the He intershell, assuming that some small-scale mixing 
occurs between the H-rich convective envelope and the $^{12}$C-rich He intershell at the 
end of each third dredge up. At H-reignition, proton captures on $^{12}$C produce $^{13}$C and a 
$^{13}$C {\it pocket} is formed. The $^{13}$C is subsequently consumed via ($\alpha,n$) 
reactions, neutrons are released and 
$s$-process isotopes are overproduced by factors of up to several thousand times their
initial abundances. The pocket is then engulfed in the convective zone driven by
the thermal instability, and diluted by a factor of $\sim$ 1/20 with material
from H-burning ashes and other He-intershell material.

During the thermal pulse a second small neutron exposure occurs, produced by
the $^{22}$Ne($\alpha,n$)$^{25}$Mg reaction, which is marginally activated at the
bottom of the convective zone when the temperature reaches above $2.5 \times
10^{8}$ K. The marginal activation of the $^{22}$Ne neutron source  
produces a small neutron burst of short duration ($\sim$ 6 yr) but with a high
average peak neutron density, $\sim 10^{11} n$ cm$^{-3}$, affecting the final 
composition of branching-dependent isotopes.
After the quenching of the thermal pulse the following third 
dredge up episode is mixes the $s$-processed material to the surface,
where it is spectroscopically observed and it is included in SiC grains.

High-precision laboratory measurements of the isotopic compositions of 
SiC grains represent the most detailed record of the composition of AGB stars, and thus a 
major constraint for the theoretical models of these stars.
In this paper we describe the technique used to analyse the composition of 
heavy elements in SiC grains (Section 2), a few main conclusions derived from comparing 
laboratory data to stellar models (Section 3) and predictions for the compositions in AGB 
envelopes of some Fe-peak elements (Section 4), which are good candidates for future 
high-precision measurements and will set further constraints to our knowledge of 
AGB stars. 

\section{The Chicago-Argonne Resonant Ionisation Spectrometer for Mass Analysis 
(CHARISMA)}

To measure trace element compositions in single grains, instruments of very high 
sensitivity are needed to detect as much as possible of the low concentration of such 
elements. Moreover, for many heavy elements, isobaric $interferences$
are present. This means that it is not possible to distinguish between stable isobars, such as 
$^{96}$Mo and $^{96}$Zr.

In the mid-1990s a new instrument of high sensitivity was developed at the Argonne 
National Laboratory with which it is possible to analyse the composition of trace
elements in single presolar grains of the relatively large size of a few $\mu$m
\citep{ma:95,savina:03b}.
This technique makes use of resonance ionisation mass spectrometry (RIMS). 
One or more lasers are tuned on the same energies needed to excite an atom
to higher and higher energy levels, i.e. the lasers are in {\it resonance} with the
atomic levels, until the energy of the atom is above its ionisation potential,
electrons are freed and an ion is created. 
The extracted ions are then separated by mass by going through a Time-Of-Flight mass 
spectrometer, which separates different masses in time by using the differences in 
transit time of ions of different masses in an electric field. 
Once the ions are separated their number is analysed using a ion detector. 
Since each element has a unique energy level 
structure, RIMS provides a ionisation method that selects which element is going to be 
ionised and hence mass interferences are automatically avoided. Since this method must 
be applied to material in the gas phase, the solid grains are first vaporized by laser 
ablation. The RIMS technique has extremely high
sensitivity so that enough ions are extracted even in the case of trace elements to
allow a relatively precise measure of the isotopic composition of the element.

This technique has been applied to date to the measurement of Zr \citep{nicolussi:97}, Mo
\citep{nicolussi:98a}, Sr \citep{nicolussi:98b}, Ba \citep{savina:03a} and Ru \citep{savina:04}
in single presolar large SiC and graphite grains \citep{nicolussi:98c} from the Murchison
meteorite. These data are of invaluable significance in the 
study of the nucleosynthetic processes that produce the elements heavier than iron.

\section{CHARISMA data and AGB stars}

\begin{figure}
\includegraphics[width=6.5cm]{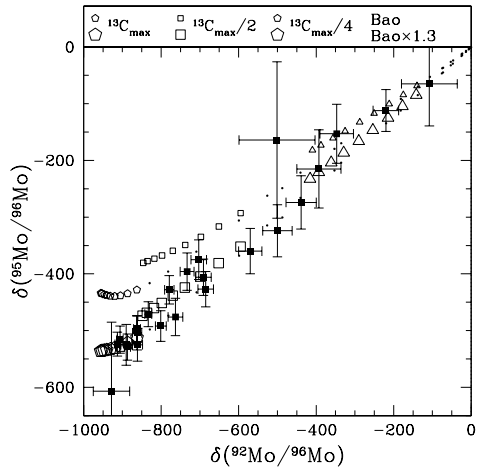}
\caption{
Plot of the $^{95}$Mo/$^{96}$Mo versus $^{92}$Mo/$^{96}$Mo ratios ($\delta$ = permil 
variation with respect to solar) measured (black squares with 2$\sigma$ errorbars) in 
single SiC and predicted (small dots and open symbols) by theoretical models at the 
surface of AGB stars of initial solar composition ($\delta$=0 \permil) and mass 1.5 $M_{\odot}$ 
for different choices of the amount of the $^{13}$C neutron source. Each symbol in the 
predictions corresponds to a third dredge up episode. Open symbols represent 
situations in which C/O $>$ 1 in the envelope, a necessary condition for the formation of SiC 
grains. 
The larger symbols for model predictions plot the same cases as the smaller symbols, 
except that the value of the $^{95}$Mo(n,$\gamma$)$^{96}$Mo reaction has been multiplied 
by a factor of 1.3 with respect to the current recommended value.
}
\label{fig:Mo}
\end{figure}

The implications for AGB models of data for Zr, Mo, Sr and Ba in single SiC have been
studied in detail by \citet{lugaro:03}. One interesting example is the composition of
$^{92}$Mo, $^{95}$Mo and $^{96}$Mo shown in Figure 1 in form of a three-isotope plot in
which two isotopic ratios, $^{95}$Mo/$^{96}$Mo and $^{92}$Mo/$^{96}$Mo in this case, with
a common reference isotope, are plotted as function of one another. Three-isotope
plots are widely used when comparing isotopic ratios because they have the helpful
property that a composition resulting from the mixing of two types of $components$ lies on a 
straight line connecting the points representing the two components.  
In the case of Figure \ref{fig:Mo}, the data and the predicted values lay on a straight line 
connecting the component of solar composition ($\delta$=0 \permil) with the $s$-process 
component from the He intershell. Because $^{92}$Mo is a $p$-only isotope, while 
$^{96}$Mo is a $s$-only isotope $\delta$($^{92}$Mo/$^{96}$Mo)$_s = - 1000$ \permil\
for the $s$-process component. Because $\sigma_n(A) N(A) \sim constant$ (where $\sigma_n$(A) 
is the neutron capture cross section, and $N(A)$ the abundance in number of a given 
isotope $A$) locally during the $s$ process, ($^{95}$Mo/$^{96}$Mo)$_s$ $\simeq$ 
$\sigma_n$($^{96}$Mo)/$\sigma_n$($^{95}$Mo).
Thus the slope of $\delta$($^{95}$Mo/$^{96}$Mo) versus $\delta$($^{92}$Mo/$^{96}$Mo) is an 
indication of the value of the neutron capture cross 
sections of these isotopes. 
The precision with which presolar grain data are obtained opens opportunities in 
predicting values of neutron capture cross sections thus stimulating new nuclear 
experiments. For example, SiC data for Mo isotopes are best matched when
increasing by 30\% the value of the $^{95}$Mo(n,$\gamma$)$^{96}$Mo reaction recommended
by \citet{bao:00}. Cross sections of Mo, and Zr, isotopes are known with a considerable 
uncertainty and more precise measurements are highly desirable.

When analysing the Zr composition of single SiC \citep[see Figure 5 of][]{lugaro:03}, it is easy 
to see that the mixing lines in the three isotope plots are far from being straight lines. This 
means that the $s$-process component from the He intershell of AGB star is not constant. A 
wealth of precise of information can be thus derived on 
AGB models. First, $^{90,91,92}$Zr have or are close to have magic number of 
neutrons. Their neutron capture cross sections are very low, 
so their production is very sensitive to the main neutron exposure in the $^{13}$C pocket. The 
spread in the $^{90,91,92}$Zr/$^{94}$Zr ratio observed in single presolar SiC grains is matched 
only by allowing a spread of efficiencies in the neutron flux in the $^{13}$C pocket.
On the other hand, $^{96}$Zr is produced via a branching at the unstable $^{95}$Zr only 
if the neutron density exceeds $\simeq$ 5 $\times 10^8$ neutrons/cm$^3$. Data from 
single SiC grains show deficits in the $^{96}$Zr/$^{94}$Zr ratio with respect to solar
and point to a marginal activation of $^{22}$Ne neutron source in the grain parent stars.

\section{A future opportunity: the Fe-peak elements}

The Galactic production of elements belonging to the Fe peak is due to the operation of 
the $equilibrium$ process in supernovae. These elements are not produced by the $s$ 
process, however, their isotopic composition in AGB stars can be modified by neutron 
captures. Because of their high solar abundances and low neutron capture cross sections, 
modifications to the compositions of Fe-peak isotopes in AGB stars are mostly 
due to the activation of the $^{22}$Ne neutron source in the convective thermal pulse.

In Figure 2, 3 and 4 we present the predicted composition of Cr, Fe and Ni isotopes at 
the surface of AGB models of 1.5 and 3 \msun\ and solar metallicity for different choices 
of the $^{13}$C amount in the $^{13}$C pocket. As in Figure \ref{fig:Mo}, 
$\delta$ represent permil variation with respect to solar and open symbols are used
when C/O$>$1. The points representing the final composition for each $^{13}$C pocket case of a 
given mass are connected by a thick solid line.

   \begin{figure*}
   \centering
   \resizebox{\hsize}{!}{\includegraphics{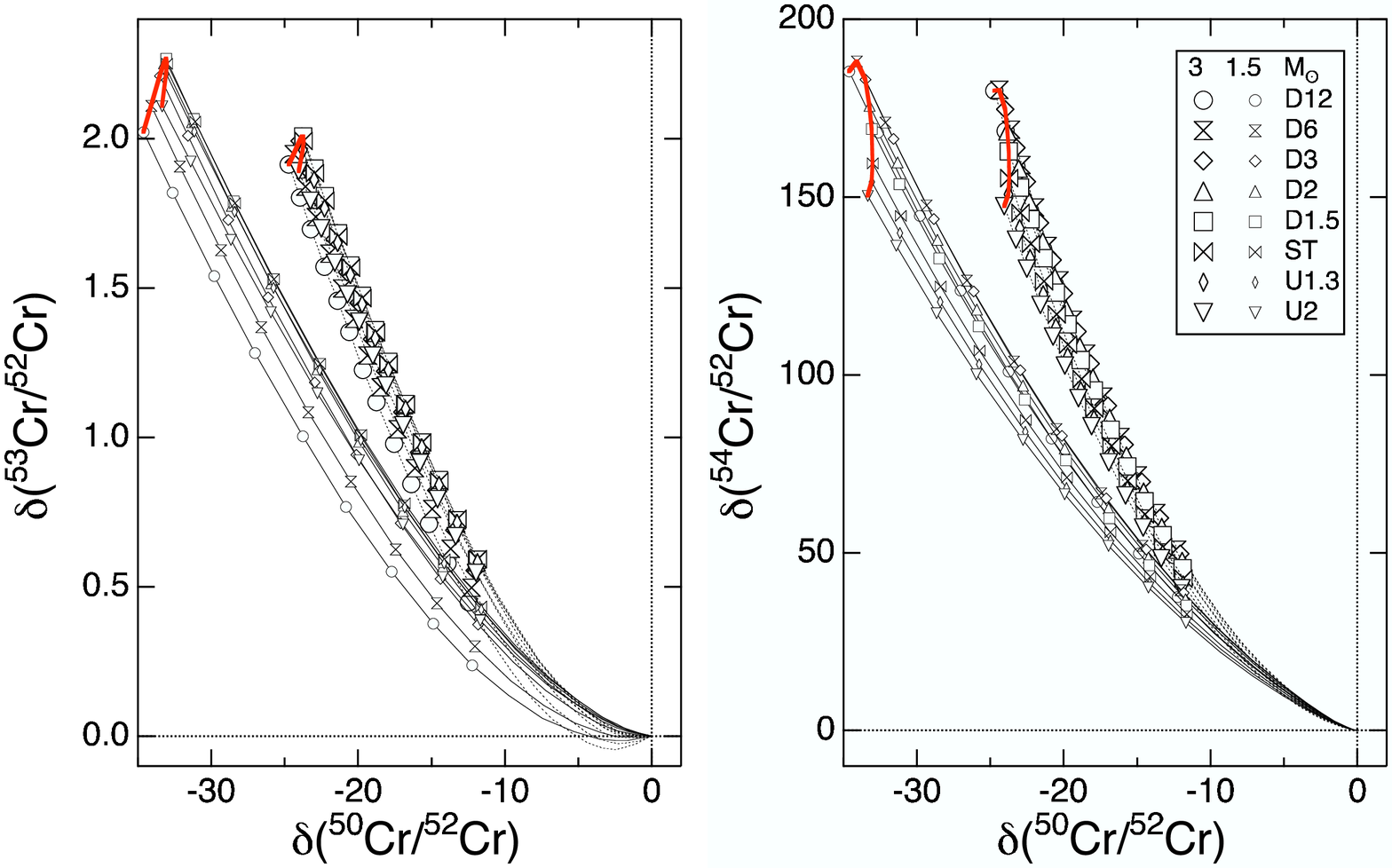}}
   \caption{Predicted isotopic composition of Cr.}
              \label{fig:Cr}%
    \end{figure*}

   \begin{figure*}
   \centering
   \resizebox{\hsize}{!}{\includegraphics{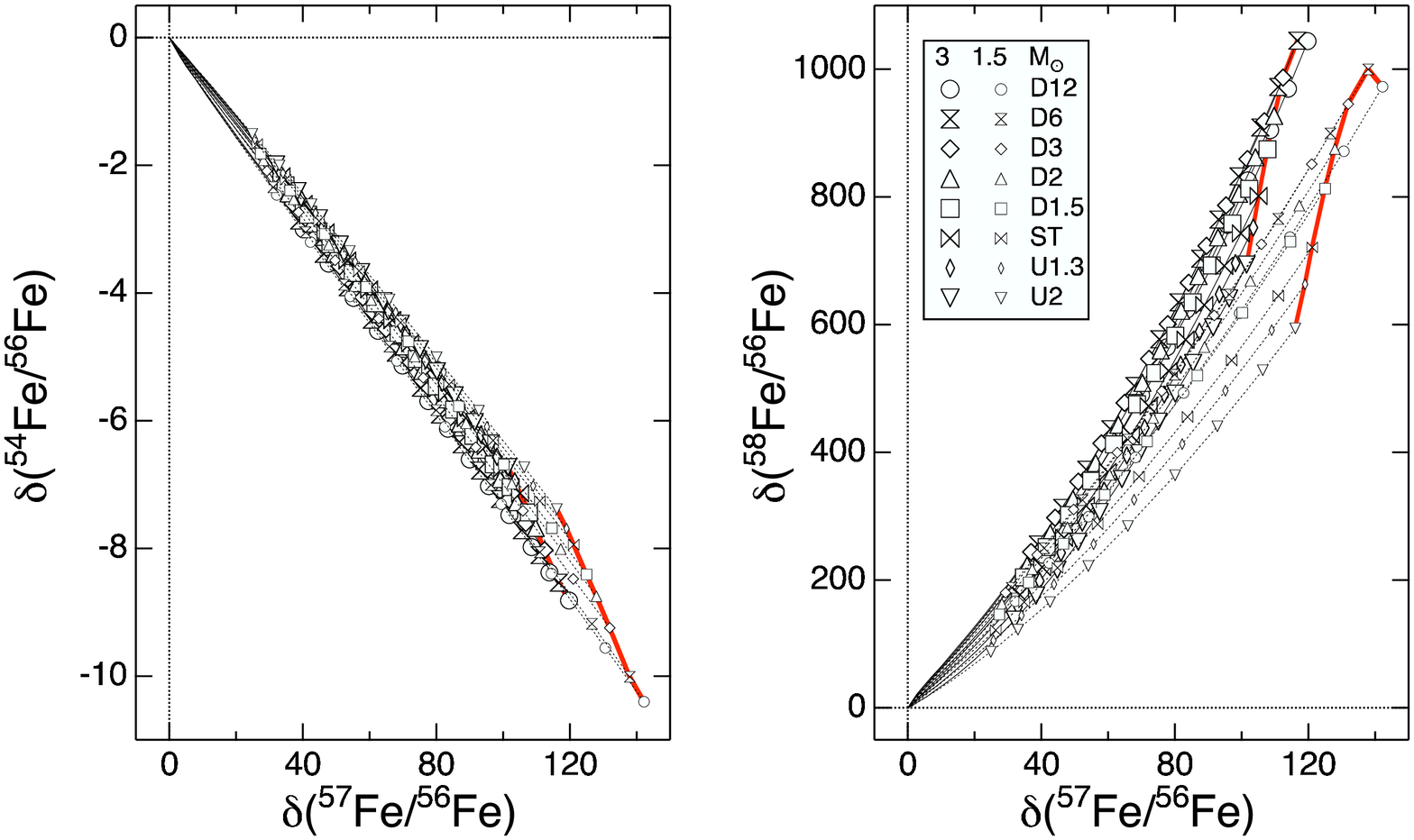}}
   \caption{Predicted isotopic composition of Fe.}
              \label{fig:Fe}%
    \end{figure*}

   \begin{figure*}
   \centering
   \resizebox{\hsize}{!}{\includegraphics{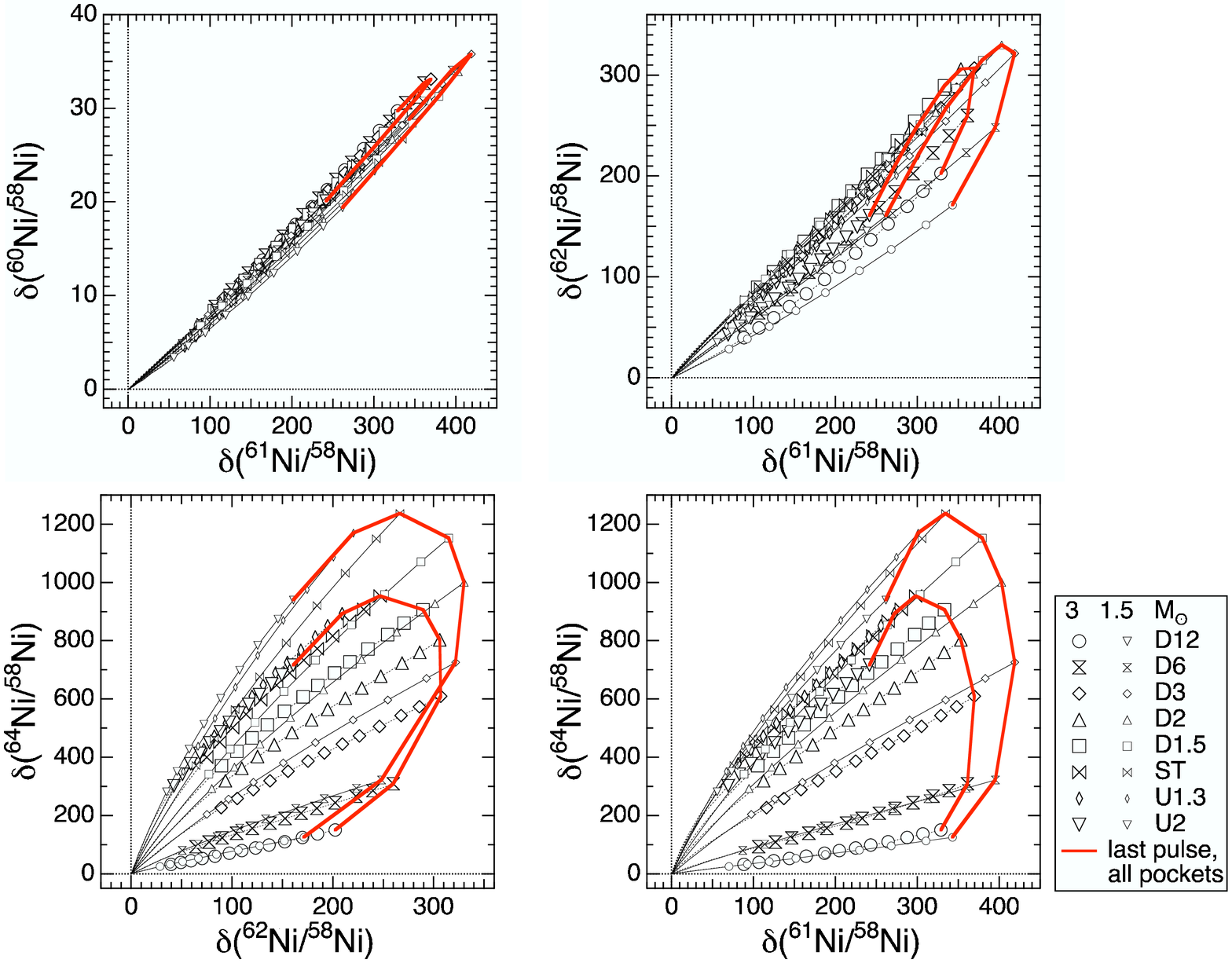}}
   \caption{Predicted isotopic composition of Ni.}
              \label{fig:Ni}%
    \end{figure*}

As for the Cr isotopic composition, only the $^{54}$Cr/$^{52}$Cr ratio is significantly modified 
by nucleosynthesis in AGB stars, increasing of up to 20\% of its solar value. 

Among the Fe isotopic ratios, $^{58}$Fe/$^{56}$Fe stands out reaching up to 
twice the solar value. Because of the very low abundance of $^{58}$Fe (0.28\% of Fe in the 
solar system) it is a challenge to measure this isotopic ratio in SiC. If this task is 
accomplished, strong constraints will be set on the operation of the $^{22}$Ne neutron source 
in AGB stars, which is mostly responsible for the increase of the $^{58}$Fe/$^{56}$Fe 
ratio. 

Another isotopic ratio of much interest is represented by the $^{64}$Ni/$^{58}$Ni ratio. 
As shown in Figure \ref{fig:Ni}, unlike the other Fe-peak isotopic ratios presented here, the 
$^{64}$Ni/$^{58}$Ni ratio varies greatly with the choice of the $^{13}$C amount in the pocket, 
ranging from 20\% to more than factor of two of its solar value. Measurements of this ratio
will represent further indications of the features of the neutron flux in the $^{13}$C pocket, in 
particular they could confirm, or maybe not, the fact that a spread of $s$-process efficiencies 
occurs in the $^{13}$C pocket. We also note that the composition of Fe-peak element depends on 
the choice of mass loss used to compute the AGB models and thus could represent a way to better 
constrain this uncertain parameter of AGB evolution.

The composition of Fe-peak elements can also be affected by the 
initial composition of the star and thus measurements of such composition could set 
detailed constraints on the Galactic chemical evolution of isotopes of these elements.

As measurement techniques improve fast with modern technology the opportunities coming from the 
study of presolar grains are compelling and cannot be ignored in the future of AGB modelling. 

\begin{acknowledgements}

This research was supported by the IoA PPARC rolling grant, the National Aeronautics and Space 
Administration, the Italian MIUR-FIRB Project ``Astrophysical origin 
of the heavy elements beyond iron'' and the Department of Energy, BES-Material 
Sciences through Contract No. W-31-109-ENG-38.

\end{acknowledgements}

\bibliographystyle{aa}

\end{document}